\documentclass[11pt,twoside]{article}
\usepackage{asp2010}
\usepackage{hyperref}

\resetcounters

\begin{document}

\title{The Galactic O-Star Spectroscopic (GOSSS) and Northern Massive Dim Stars (NoMaDS) Surveys, the Galactic O-Star Catalog (GOSC),
and Marxist Ghost Buster (MGB)}
\author{J. Ma\'{\i}z Apell\'aniz$^1$, A. Pellerin$^{2,3}$, R. H. Barb\'a$^{4,5}$, S. Sim\'on-D\'{\i}az$^{6,7}$, E. J. Alfaro$^1$, N. I. Morrell$^8$, 
A. Sota$^1$, M. Penad\'es Ordaz$^1$, and A. T. Gallego Calvente$^1$
\affil{$^1$Instituto de Astrof\'{\i}sica de Andaluc\'{\i}a-CSIC, E-18008 Granada, Spain}
\affil{$^2$Department of Physics and Astronomy, Texas A\&M University, College Station, TX 77845, USA}
\affil{$^3$George P. and Cynthia Woods Mitchell Institute for Fundamental Physics and Astronomy, Texas A\&M University, 
College Station, TX 77845, USA}
\affil{$^4$Departamento de F\'{\i}sica, Universidad de La Serena, La Serena, Chile}
\affil{$^5$Inst. de CC. Astron\'omicas, la Tierra y el Espacio, 5400 San Juan, Argentina}
\affil{$^6$Instituto de Astrof\'{\i}sica de Canarias, E-38205 La Laguna (Tenerife), Spain}
\affil{$^7$Dep. de Astrof\'{\i}sica, Un. de La Laguna, E-38205 La Laguna (Tenerife), Spain}
\affil{$^8$Observatories of the Carnegie Institution of Washington, La Serena, Chile}}

\begin{abstract}
There are several ongoing massive-star (mostly of spectral type O) surveys that are significantly increasing the quality and quantity of
the spectroscopic information about these objects. Here we discuss and present results for two of them, GOSSS and NoMaDS. We also discuss
recent and future developments on the Galactic O-Star Catalog and announce the upcoming availability of Marxist Ghost Buster, an IDL code
that attacks spectral classification (hence the name) by using an interactive comparison with spectral libraries. 
\end{abstract}

\section{O-star spectroscopic surveys}

$\,\!$\indent The last few years have seen a boom in spectroscopic surveys of O stars, a consequence of the availability of modern
instrumentation at telescopes of different apertures (from $\sim$1 m to $\sim$10 m) and the realization of the importance of high-quality
data and large samples to characterize massive stars. These surveys differ from those in previous decades by their sample sizes, high
S/N, and, in some cases, superior spectral resolving power and multi-epoch nature. Their usefulness is also helped by companion high-resolution 
imaging surveys to better characterize multiplicity (e.g. \citealt{Maiz10}), the development of dedicated reduction pipelines, and, 
by the public availability of the reduced spectra. Table~\ref{Osurveys} lists the characteristics of some of those surveys. All of them 
include the blue-violet region of the spectrum traditionally used for spectral classification but some extend that towards the rest of the
visible wavelengths.

\begin{table}
\caption{Current spectroscopic surveys with large samples of O stars}
\label{Osurveys}
\centerline{
\footnotesize
\begin{tabular}{@{\extracolsep{0pt}}llll} 
 \\
\hline
\multicolumn{1}{c}{Acronym}         &
\multicolumn{1}{c}{Name}            &
\multicolumn{1}{c}{Principal}       &
\multicolumn{1}{c}{Reference}       \\
                                    &
                                    &
\multicolumn{1}{c}{Investigator(s)} &
                                    \\
\hline
GOSSS    & Galactic O-Star Spectroscopic Survey  & J. Ma\'{\i}z Apell\'aniz & \citet{Maizetal10b}          \\
OWN      & Spectroscopic survey of Galactic O    & R. H. Barb\'a,           & \citet{Barbetal10}           \\
         & $\;\;$and WN stars                    & $\;\;$N. I. Morrell, \&  &                              \\
         &                                       & $\;\;$R. C. Gamen        &                              \\
IACOB    & IACOB spectroscopic survey            & S. Sim\'on-D\'{\i}az     & \citet{SimDetal10}           \\
NoMaDS   & Northern Massive Dim Stars survey     & A. Pellerin              & This contribution            \\
VFTS     & VLT-FLAMES Tarantula Survey           & C. Evans                 & \citet{Evanetal10}           \\
RIOTS4   & Runaways and Isolated O-Type Star     & M. S. Oey \&             & \citet{OeyLamb11}            \\
         & $\;\;$Spectroscopic Survey of the SMC & $\;\;$J. B. Lamb         &                              \\
Gaia-ESO & Gaia-ESO spectroscopic survey         & G. Gilmore \&            & \url{http://www.gaia-eso.eu} \\
         &                                       & $\;\;$S. Randich         &                              \\
\hline
\end{tabular}
}
\centerline{
\footnotesize
\begin{tabular}{@{\extracolsep{0pt}}lrrrllrr} 
\hline
\multicolumn{1}{c}{Acronym}         &
\multicolumn{1}{c}{\# of}           &
\multicolumn{1}{c}{\# of}           &
\multicolumn{1}{c}{Resol.}          &
\multicolumn{1}{c}{Sample}          &
\multicolumn{1}{c}{Spectral}        &
\multicolumn{1}{c}{\# of}           &
\multicolumn{1}{c}{Fract.}          \\
                                    &
\multicolumn{1}{c}{targets}         &
\multicolumn{1}{c}{O stars}         &
                                    &
                                    &
\multicolumn{1}{c}{types}           &
\multicolumn{1}{c}{epochs}          &
\multicolumn{1}{c}{compl.}          \\
\hline
GOSSS    &      3000 &      1500 &      2500 & Galactic N + S    & O (+ B + WR + \ldots) &  1-6 &  45\% \\
OWN      &       300 &       250 &   15\,000 & Galactic S        & O + WN                &   10 &  70\% \\
         &           &           & - 48\,000 &                   &                       &      &       \\
IACOB    &       300 &       150 &   23\,000 & Galactic N        & O + B                 &  1-5 &  80\% \\
         &           &           & - 46\,000 &                   &                       &      &       \\
NoMaDS   &       200 &       150 &   30\,000 & Galactic N        & O (+ B + WR)          &  1-5 &  20\% \\
VFTS     &      1000 &       250 &      7000 & 30 Doradus        & O + B + \ldots        &    9 & 100\% \\
RIOTS4   &       374 & $\sim$130 &      2600 & SMC field         & O + B + WR            & 1-12 &  90\% \\
Gaia-ESO &  100\,000 & $\sim$100 &   20\,000 & Galactic clusters & All                   &    1 &   0\% \\
         &           &           &           & $\;\;$+ field     &                       &      &       \\
\hline
\end{tabular}
}
\end{table}

\section{The Galactic O-Star Spectroscopic Survey (GOSSS)}

$\,\!$\indent The Galactic O-Star Spectroscopic Survey (GOSSS, \citealt{Maizetal10b}) is observing a sample of $\sim$3000 stars
initially selected from a literature search of previous spectral classifications\footnote{The search is ongoing and we keep adding stars.}.
The main criterion for inclusion in the survey is the existence of at least one previous classificaion as spectral type O. It has currently 
observed over 1300 targets and our projections indicate that $\sim$50\% of the objects in the full sample are indeed O stars; the rest are 
misclassifications due to low S/N data, incorrect classification techniques, insuficient resolution, or simple misidentifications. GOSSS has different 
quality checks to ensure that the data have good S/N (200 minimum, $>$300 in most cases) and uniform spectral resolving power in the 
3900-5100 \AA\ classification range. The GOSSS data are being obtained at one telescope in the southern hemisphere (the 2.5 m du Pont at 
Las Campanas Observatory, Chile) and three telescopes in the northern hemisphere (the 1.5 m telescope at the Observatorio de Sierra Nevada,
the 3.5 m telescope at Calar Alto, and the 4.2 m William Herschel Telescope at La Palma, all of them in Spain).

	Compared to the other surveys in Table~\ref{Osurveys}, GOSSS has two main differences: its final sample of O stars will be larger
by an order of magnitude but its spectral resolving power will be significantly lower (with the exception of RIOTS4). Those are a consequence of the 
different philosophies behind the surveys. GOSSS aims to detect a large fraction of the O stars in the solar neighborhood by analyzing all good
candidates down to a given magnitude limit ($B\sim 13$). A spectral resolving power of 2500 is sufficient to obtain accurate spectral types
(and thus determine which ones are real O stars) but not to determine all physical parameters of interest such as $v\sin i$, detect SB2s with maximum 
velocity separations lower than $\sim$150 km/s, or to study
the kinematics of the intervening ISM. The latter properties can be obtained with the higher resolution surveys, which are
restricted to smaller samples of brighter stars (OWN, IACOB) or require more expensive time in larger telescopes (NoMaDS). Those larger
telescopes are in some cases only accessible to large collaborations that target clusters with multi-fiber spectrographs (VFTS, Gaia-ESO);
those programs can have large samples but only a minority of them are O stars. In summary, GOSSS excels at giving us better statistics and
searching for interesting targets that can later be analyzed in more detail at higher resolution with the other surveys.

	The first GOSSS results were published in a letter \citep{Walbetal10a} that introduced the new class of Ofc stars and presented two
new examples of the rare Of?p class, NGC 1624-2 and CPD -28$^{\rm o}$ 2561. The first main block of the survey (northern bright stars) was
later published \citep{Sotaetal11}. A second letter \citep{Walbetal11} about nitrogen-rich fast rotators is scheduled to appear later in
2011. A second block of the survey (southern bright stars) will be submitted later this year.

\section{The Northern Massive Dim Stars (NoMaDS) survey}

$\,\!$\indent OWN (southern hemisphere) and IACOB (northern hemisphere) are the high-resolution counterparts of GOSSS in the sense of attempting
to obtain spectra for all Galactic O stars (plus other massive stars) down to a given magnitude limit. OWN originally had an emphasis on the detection of
SB2s and IACOB on the detailed modeling of physical parameters; later on, both surveys started a collaboration given their complementarity in coordinate
coverage and the possibility of using the other one's data for their original purposes. However, given that those surveys use 2-m
class telescopes and that they aim to obtain similar S/N (per resolution element) values with a spectral resolving power more than an order of
magnitude higher, they are limited to a significantly brighter magnitude limit ($B=8-9$). That leaves outside their sample many interesting
targets, such as NGC 1624-2 or those in the Cyg OB2 association. 

	In order to remedy that situation, we have started NoMaDS, a survey that complements OWN and IACOB by obtaining $R=30\,000$ optical
spectra of northern massive stars down to $B=14$. NoMaDS data are being obtained with the High Resolution Spectrograph at the Hobby-Eberly
Telescope (HET), located at McDonald Observatory, Texas, USA. The large effective aperture of HET (9.2 m) allows the obtention of
high-resolution spectra with S/N of 200-300 for stars much dimmer than those in OWN or IACOB with the same exposure time.

	We have obtained spectra with HET for 40 stars in the first four months of the survey. In most cases we observed only a single epoch but 
for a few targets we observed up to four epochs. Among our first results, we have obtained the first high-resolution spectra ever of NGC 1624-2, 
the most extreme of the currently known Of?p stars. We have also followed the orbit of Cyg OB2-9 near its periastron passage of June 2011 and we 
have obtained separate spectral types for both components. We plan to continue NoMaDS during 2011 and 2012 until we obtain 250 spectra for 200 
stars.

\section{The Galactic O-Star Catalog (GOSC)}

$\,\!$\indent The first published version (v1.0) of the Galactic O-Star Catalog (GOSC) originated with the 378-star sample of \citet{Maizetal04b}, 
which was compiled from a collection of accurate spectral types from the literature (the vast majority of them coming from Nolan Walborn). 
This was later extended in v2.0 by \citet{Sotaetal08} to 
include 1208 objects, some of them not O stars, with spectral types of diverse origins and qualities. Subsequent small revisions extended the 
number of objects to 1285 in the last public version (v2.3.2, April 2011). 

	While analyzing the GOSSS spectra we realized that a significant fraction of the stars in v2.0 originally classified as being of O
type were actually early-B stars or of even later type. Also, while writing \citet{Sotaetal11} we realized the need to introduce the 
spectral subtype O9.7 for luminosity classes V to III and to consequently tweak the spectral classification criteria for late-O and early-B
stars. Those two issues led us to alter our policy regarding the addition of new stars to GOSC: We stopped adding new stars to the public
version and instead created a private version that currently has 3000+ additional objects\footnote{Some are known to be massive non-O stars, 
others are unobservable by GOSSS because they are too dim in the $B$ band or because they are not easily resolved from the ground.} and 
is used to generate the GOSSS sample. By keeping that version private our intention is to reduce the noise (in the form of poor-quality spectral
classifications) until we can increase the signal. We will start doing that when we roll out v3.0 of GOSC in 2012 after we publish the second block
of GOSSS ($\sim$200 bright southern stars). 
Version 3.0 will replace the old spectral types with the ones derived from GOSSS and will include the results from the first two GOSSS
blocks. Additional spectral types will be added to GOSC when subsequent papers are published. 

	Other recent (last two years) and future changes to GOSC include:

\begin{itemize} 
\setlength{\itemsep}{0pt}
  \item A simplified URL: \url{http://gosc.iaa.es}.
  \item A new interface that allows for searches and filters using a combination of IDL, Javascript, and MySQL.
  \item The possibility of selecting HTML, Aladin, and VOTable outputs.
  \item In the future we will use GOSC as the platform to disseminate the data from our surveys, including GOSSS, NoMaDS, and our AstraLux-based
        imaging surveys \citep{Maiz10}.
\end{itemize}

\section{Marxist Ghost Buster (MGB)}

$\,\!$\indent The first author (J. M. A.) has written an IDL code called Marxist Ghost Buster (or MGB) as a companion to the GOSSS (and other) 
surveys in order to attack the spectral classification of the data. The code derives its name from attacking both classes (hence the Marxist part) 
and spectra\footnote{In the Latin meaning(s) of the word.} (hence the Ghost Buster part). 
The spectral classification is done by comparing the observed spectra with a standard library. The current
beta version includes two libraries, both intended for classification of OB stars in the blue-violet region of the spectrum. The first library is
built from GOSSS data at $R\sim 2500$ and is being used for the spectral classification of GOSSS itself. The second library has been compiled by Sergio 
Sim\'on-D\'{\i}az (northern stars, from the IACOB survey) and Hughes Sana (southern stars, from a variety of sources), has a higher spectral resolving power 
($R\sim 4000$), and is being used for the spectral classification of a subset of the VFTS data.
In any case, MGB is a general code that can be adapted for the spectral classification of other types of stars as long as
a library is generated.

	MGB is an interactive code where the user is allowed to change four parameters:

\begin{enumerate}
\setlength{\itemsep}{0pt}
  \item Spectral subtype.
  \item Luminosity class.
  \item $n$ index (broadening).
  \item Different standards for the same grid point in order to evaluate aspects such as ONC or $f$ characteristics.
\end{enumerate}

	Additionally, it is possible to build synthetic binaries with different velocities and component types in order to fit SB2 spectra.

	MGB is currently being tested. Its first public version will be released at the first author's web site (\url{http://jmaiz.iaa.es}) at
or near the same time v3.0 of GOSC becomes available during 2012\footnote{And no, we do not believe that is another sign of the end of
the world.}.

\hyphenation{Mi-nis-te-rio}

\acknowledgements J. M. A. is grateful to the Department of Physics and Astronomy at Texas A\&M University for their hospitality during most of
2011. Support for this work was provided by [a] the Spanish Ministerio de Ciencia e Innovaci\'on through 
grants AYA2007-64712 and AYA2010-17631, the Ram\'on y Cajal Fellowship program, and FEDER funds; [b] the Junta de Andaluc\'{\i}a
grant P08-TIC-4075; and [c] the George P. and Cynthia Woods Mitchell Institute for Fundamental Physics and Astronomy.

\bibliography{general}

\begin{thebibliography}{}
\expandafter\ifx\csname natexlab\endcsname\relax\def\natexlab#1{#1}\fi
\expandafter\ifx\csname url\endcsname\relax
  \def\url#1{\texttt{#1}}\fi
\expandafter\ifx\csname urlprefix\endcsname\relax\def\urlprefix{URL }\fi
\providecommand{\eprint}[2][]{\url{#2}}

\bibitem[{Barb{\'a} et~al.(2010)Barb{\'a}, Gamen, Arias, Morrell,
  Ma{\'{\i}}z~Apell{\'a}niz, Alfaro, Walborn, \& Sota}]{Barbetal10}
Barb{\'a}, R.~H., Gamen, R.~C., Arias, J.~I., Morrell, N.~I.,
  Ma{\'{\i}}z~Apell{\'a}niz, J., Alfaro, E.~J., Walborn, N.~R., \& Sota, A.
  2010, in Rev. Mex. Astron. Astrof{\'\i}s. (conference series), vol.~38, 30

\bibitem[{Evans et~al.(2010)}]{Evanetal10}
Evans, C.~J., et~al. 2010, in IAU Symposium, edited by {R.~de Grijs \&
  J.~R.~D.~L{\'e}pine}, vol. 266 of IAU Symposium, 35

\bibitem[{Ma{\'{\i}}z~Apell{\'a}niz(2010)}]{Maiz10}
Ma{\'{\i}}z~Apell{\'a}niz, J. 2010, A\&A, 518, A1+

\bibitem[{Ma{\'{\i}}z~Apell{\'a}niz et~al.(2010)Ma{\'{\i}}z~Apell{\'a}niz,
  Sota, Walborn, Alfaro, Barb{\'a}, Morrell, Gamen, Arias, \&
  Penad{\'e}s~Ordaz}]{Maizetal10b}
Ma{\'{\i}}z~Apell{\'a}niz, J., Sota, A., Walborn, N.~R., Alfaro, E.~J.,
  Barb{\'a}, R.~H., Morrell, N.~I., Gamen, R.~C., Arias, J.~I., \&
  Penad{\'e}s~Ordaz, M. 2010. \eprint{arXiv:1010.5680}

\bibitem[{Ma{\'\i}z~Apell{\'a}niz et~al.(2004)Ma{\'\i}z~Apell{\'a}niz, Walborn,
  Galu{\'e}, \& Wei}]{Maizetal04b}
Ma{\'\i}z~Apell{\'a}niz, J., Walborn, N.~R., Galu{\'e}, H.~{\'A}., \& Wei,
  L.~H. 2004, ApJS, 151, 103

\bibitem[{Oey \& Lamb(2011)}]{OeyLamb11}
Oey, M.~S., \& Lamb, J.~B. 2011. \eprint{arXiv:1109.0759}

\bibitem[{Sim{\'o}n-D{\'\i}az et~al.(2010)Sim{\'o}n-D{\'\i}az, Garc{\'\i}a,
  Castro, \& Herrero}]{SimDetal10}
Sim{\'o}n-D{\'\i}az, S., Garc{\'\i}a, M., Castro, N., \& Herrero, A. 2010.
  \eprint{arXiv:1009.3750}

\bibitem[{Sota et~al.(2011)Sota, Ma{\'{\i}}z~Apell{\'a}niz, Walborn, Alfaro,
  Barb{\'a}, Morrell, Gamen, \& Arias}]{Sotaetal11}
Sota, A., Ma{\'{\i}}z~Apell{\'a}niz, J., Walborn, N.~R., Alfaro, E.~J.,
  Barb{\'a}, R.~H., Morrell, N.~I., Gamen, R.~C., \& Arias, J.~I. 2011, ApJS,
  193, 24

\bibitem[{Sota et~al.(2008)Sota, Ma{\'{\i}}z~Apell{\'a}niz, Walborn, \&
  Shida}]{Sotaetal08}
Sota, A., Ma{\'{\i}}z~Apell{\'a}niz, J., Walborn, N.~R., \& Shida, R.~Y. 2008,
  in Rev. Mex. Astron. Astrof{\'\i}s. (conference series), vol.~33, 56

\bibitem[{Walborn et~al.(2011)Walborn, Ma{\'{\i}}z~Apell{\'a}niz, Sota, Alfaro,
  Morrell, Barb{\'a}, Arias, \& Gamen}]{Walbetal11}
Walborn, N.~R., Ma{\'{\i}}z~Apell{\'a}niz, J., Sota, A., Alfaro, E.~J.,
  Morrell, N.~I., Barb{\'a}, R.~H., Arias, J.~I., \& Gamen, R.~C. 2011, AJ,
  accepted. \eprint{arXiv:1109.0515}

\bibitem[{Walborn et~al.(2010)Walborn, Sota, Ma{\'{\i}}z~Apell{\'a}niz, Alfaro,
  Morrell, Barb{\'a}, Arias, \& Gamen}]{Walbetal10a}
Walborn, N.~R., Sota, A., Ma{\'{\i}}z~Apell{\'a}niz, J., Alfaro, E.~J.,
  Morrell, N.~I., Barb{\'a}, R.~H., Arias, J.~I., \& Gamen, R.~C. 2010, ApJL,
  711, L143

\end{thebibliography}

\end{document}